\documentclass{sigchi}

\usepackage{balance}  
\usepackage{graphics} 
\usepackage{times}    
\usepackage{url}      

\makeatletter
\def\url@leostyle{%
  \@ifundefined{selectfont}{\def\UrlFont{\sf}}{\def\UrlFont{\small\bf\ttfamily}}}
\makeatother
\def\pprw{8.5in}
\def\pprh{11in}

\usepackage[pdftex]{hyperref}
\hypersetup{
pdftitle={SIGCHI Conference Proceedings Format},
pdfauthor={LaTeX},
pdfkeywords={SIGCHI, proceedings, archival format},
bookmarksnumbered,
pdfstartview={FitH},
colorlinks,
citecolor=black,
filecolor=black,
linkcolor=black,
urlcolor=black,
breaklinks=true,
}

\usepackage{color}
\definecolor{orange}{RGB}{255,127,0}
\definecolor{limegreen}{RGB}{50, 205, 50}
\definecolor{violet}{RGB}{148,0,211}

\newif\ifCOMMENTS
\COMMENTStrue

\ifCOMMENTS

\else

\fi

\usepackage{multicol}
\usepackage{cuted}
\usepackage[font=small,labelfont=bf,tableposition=top]{caption}

\usepackage{algorithm}
\usepackage{algpseudocode}

\usepackage{booktabs}

\begin{document}

\title{Autocomplete Textures for 3D Printing}

\numberofauthors{1}
\author{
  \alignauthor{
    Ryo Suzuki$^1$, \
    Tom Yeh$^1$, \
    Koji Yatani$^2$, \
    Mark D. Gross$^1$ \\
    \affaddr{
        $^1$University of Colorado Boulder, \
        $^2$University of Tokyo} \\
    \affaddr{%
      \{ryo.suzuki, \ tom.yeh\}@colorado.edu, \
      koji@iis-lab.org, \
      mdgross@colorado.edu
    }
  }
}


\teaser{
\includegraphics[width=2.1\columnwidth]{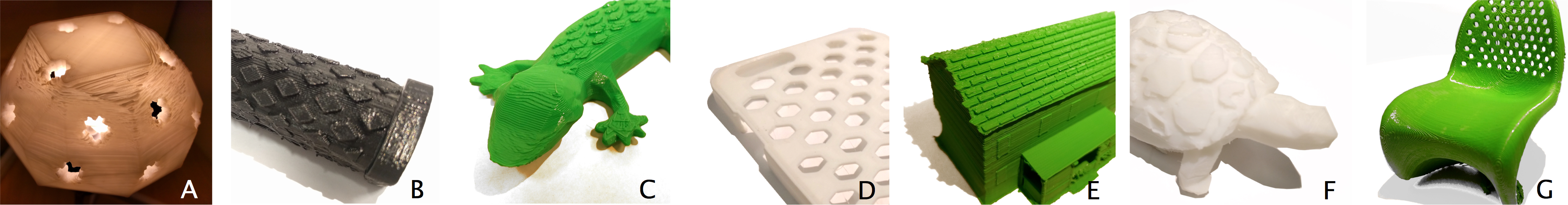}
\caption{Examples of 3D printed texture with Tabby. (A) patterns in a lampshade, (B) grips of a bike handle, (C) scales of a gecko, (D) pattern on an iPhone case, (E) a tile of an architectural model, (F) patterns of a turtle shell, and (G) aesthetic pattern on a back of a chair. In these examples, texture patterns can improve aesthetics (e.g. A, D, G), usability (e.g. B, D), and the visual details of the model (e.g. C, E, F).}~\label{fig:cover}
}

\maketitle

\begin{abstract}
Texture is an essential property of physical objects that affects aesthetics, usability, and functionality. However, designing and applying textures to 3D objects with existing tools remains difficult and time-consuming; it requires proficient 3D modeling skills. To address this, we investigated an auto-completion approach for efficient texture creation that automates the tedious, repetitive process of applying texture while allowing flexible customization. We developed techniques for users to select a target surface, sketch and manipulate a texture with 2D drawings, and then generate 3D printable textures onto an arbitrary curved surface. In a controlled experiment our tool sped texture creation by 80\% over conventional tools, a performance gain that is higher with more complex target surfaces. This result confirms that auto-completion is powerful for creating 3D textures.
\end{abstract}


\section{Introduction}

Texture is an essential property of physical objects~\cite{ashby2013materials}. 
It contributes to both aesthetics and usability~\cite{dumas2015example,gatto1978exploring,kreifeldt2011importance}.
For example, patterns in a lampshade can enhance the aesthetics of the light design. Brick patterns in architectural models can improve visual details.
Texture can also enhance ergonomic and mechanical functionality~\cite{gibson1966senses,perez2015design,torres2015hapticprint}: 
a rough texture can improve an object's usability by providing a gripping surface.
Variations in texture can offer bending and stacking functionality (e.g., living-hinges and patterns seen in Lego blocks).
Texture can also support navigation and communication for people with visual impairments~\cite{brown2012viztouch,gotzelmann2014towards,stangl2015transcribing}.

Textures are used to be considered as a feature of the material~\cite{ashby2013materials}.
It is mostly a bi-product or side effect of the choice of the material, rather than the design choice.
However, with 3D printing, texture becomes an important aspect that characterizes a printed object.
Recent 3D printing technologies enable designers to determine and assess their texture design for physical objects. 
This can open up a new opportunity for user-driven 3D object design explorations.

However, designing textures in 3D modelling is generally a tedious task for designers. 
Many textures involve a repeating pattern. 
Designers often repeat copy-and-paste to create patterns on simple flat surfaces, but this severely limits flexibility, requiring tedious manual edits to explore alternatives.
Applying texture to complex curved surfaces is difficult even for experienced designers. 
Advanced CAD tools support parametric modeling, which allows flexible customization by automating repetitive changes.
Recent work extends this approach by synthesizing textures from pre-defined examples~\cite{torres2015hapticprint}.
However, this interaction model has a significant drawback:
Parametric design inherently requires repeated parameter tuning to create and edit a model.
Such indirect manipulation significantly interrupts workflow~\cite{kazi2012vignette} and is often perceived as a barrier to inexperienced designers~\cite{norman1986user}.

\begin{figure}[t!]
\centering
\includegraphics[width=1\columnwidth]{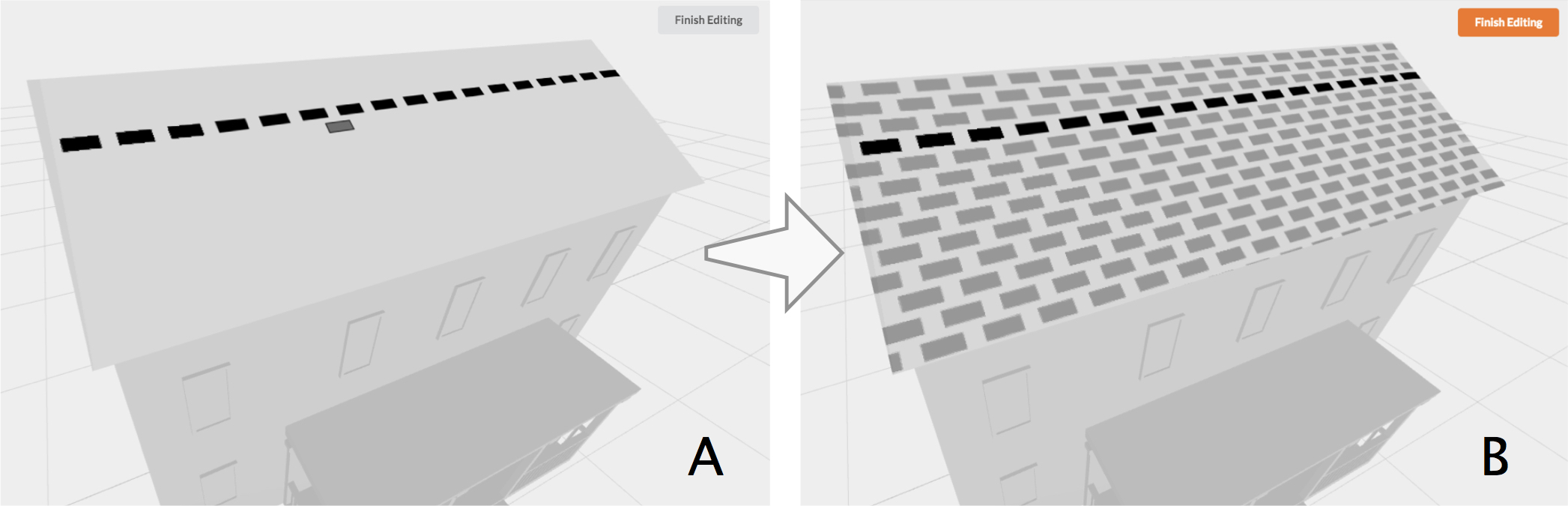}
\caption{Autocomplete textures: (A) The user interactively specifies the position of the next tactile unit. (B) Our system automatically infers the positions of other tactile units to suggest a complete texture pattern.}~\label{fig:cover-auto-completion}
\end{figure}

We present Tabby, an interactive design system that automates the repetitive operations involved in 3D texture creation while allowing the user's control over design parameters through intuitive operations.
In Tabby, a designer would only need to demonstrate the first few units of a texture pattern.
The system can automatically infer a complete pattern the designer may have in mind.
The system suggests this inferred pattern, and the designer can simply accept it or adjust to achieve the desired texture (Figure.~\ref{fig:cover-auto-completion}).

Our interaction technique is greatly inspired by recent advances in \emph{auto-completion} approaches~\cite{gulwani2011automating,xing2014autocomplete}.
Although the auto-completion technique has shown success in 2D drawing applications~\cite{kazi2014draco,xing2014autocomplete,xing2015autocomplete}, there are still challenges when applying to the 3D modeling application.
First, positioning texture elements in 3D space requires more cognitive efforts comparing to 2D drawing~\cite{hudson2016understanding}.
Our formative study shows that novice users often struggle to locate the texture by continuously changing the camera angle to ensure that texture elements were correctly attached on the surface.
Second, selecting the target region is another tedious interaction for users.
Unlike 2D drawing, specifying the target region is not a trivial problem in 3D modeling. 

We developed a series of techniques for auto-complete functionality to support 3D texture design.
To reduce the cognitive efforts to manipulate in 3D space, Tabby exploits 2D operations to design and align texture.
The system allows the user to sketch and arrange a texture element as a 2D drawing.
After designing texture patterns, the system automatically generates a 3D printable texture.
The system also infers which surface region to fill the repeated pattern.
We leverage the existing segmentation algorithms to enable automatically infer the semantic region.

We evaluate the efficiency and flexibility of designing with Tabby in a controlled experiment with seven designers.
Our study shows that Tabby speeds texture creation by 80\% over conventional tools.
This performance gain becomes even larger with more complex target surfaces. 
Our qualitative result confirms that designing and applying textures with our system is more simple and effective.

In summary, this paper includes the following three contributions to 3D object design and interactive fabrication: \begin{enumerate}
\item {Tabby, an interactive system that instantiates the auto-completion method in the context of 3D texture design;}
\item {A series of techniques to exploit 2D operation for designing 3D-printable texture patterns; and} 
\item {A controlled experiment with seven designers that show how users can efficiently and flexibly design textures comparing to conventional tools.}
\end{enumerate}

\begin{figure*}[t!]
\centering
\includegraphics[width=2.1\columnwidth]{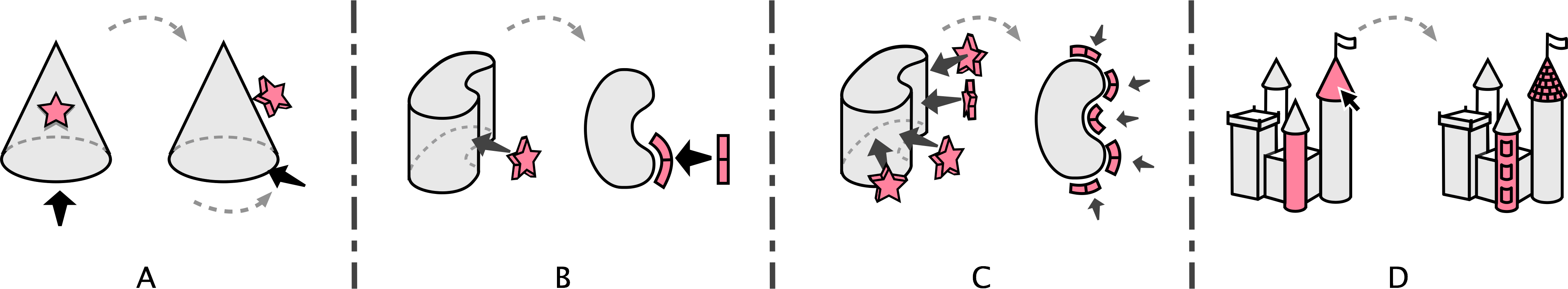}
  \caption{Four difficulties in applying texture to arbitrary surfaces: (A) Users often fail to align and manipulate 3D objects. (B) Texture surface must be deformed for non-flat surfaces. (C) Repetitive manipulation is required to adjust rotation and positioning for each element. (D) Users must perform tedious operations to choose a specific region.}~\label{fig:design-goals}
\end{figure*}

\section{Related Work}

\subsection{Geometric Texture Synthesis}
Computer graphics research has a long history of 2D image mapping~\cite{heckbert1986survey} and by-example image texture synthesis~\cite{wei2009state}.
Work on solid texture synthesis has extended 2D image synthesis to 3D models~\cite{pietroni2010solid}.
While these techniques can render and simulate 3D textures, they only generate colors or shadow on the surface, and do not change the underlying surface geometry.
Recent work has developed 3D geometric texture synthesis methods by performing voxelization \cite{bhat2004geometric}.
The synthesis process can be improved through a distance field matching~\cite{lagae2005geometry} or by restricting synthesis to only surface voxels~\cite{dumas2015example}.
A major limitation of this approach is to require voxelization to produce an output mesh.
The outcomes are sometimes undesirable because they lose the smoothness of a surface (similar to pixelization of a vector curve).
In addition, voxelization is in general computationally expensive and difficult to modify the result outcomes.
Therefore, this method is unsuitable for interactive design.

Several alternative approaches extend the geometric synthesis method.
They do not require voxelization \cite{zhou2006mesh}, and consider 3D printability or structural soundness of output models~\cite{dumas2015example}. 
However, these systems mainly aim to fully automate texture synthesis, and do not support interactive explorations of texture design.
One of our main objectives is to support interactive tactile texture creation for 3D models.

\subsection{Design Tools for 3D Texture Creation}
Prior research has developed interactive systems for various mesh editing tasks including surface deformation \cite{sorkine2004laplacian} and mesh composition \cite{schmidt2008sketch}.
Systems like MeshMixer \cite{schmidt2010meshmixer} and GeoBrush \cite{takayama2011geobrush} provide an interactive texture cloning tool, and users can copy a geometric feature from example models by a simple brushing operation.
However, these systems are not designed to support the creation of repeated patterns like tactile textures.
Exemplars of 3D tactile texture are typically difficult to obtain, inducing another limitation in existing texture cloning tools.
Also, a user must manually choose a region of both the source texture and the target surface.
Such a task can be time-consuming if the target region is large.
Tabby reduces efforts on region selection and enables a user to generate geometric features based on user-provided 2D images. 

Similar to our work, HapticPrint \cite{torres2015hapticprint} provides a design tool for tactile texture. 
With this tool, users can interactively explore a library of 2D texture patterns, choose a pattern, and apply it to an object.
HapticPrint uses the 2D height map to generate a texture mesh.
While this approach has benefits to create 3D tactile textures, it sometimes requires a re-meshing process to increase the vertex density of the target surface.
The capability of HapticPrint is also limited to bumpy textures, and hole or hollow patterns are not supported.
Our system supports the creation of a wider variety of 3D textures as well as flexible design explorations through auto-complete approaches.

Landreneau et al. \cite{landreneau2010scales} developed an interactive tool for scale pattern synthesis.
In the workflow, the user can fill a region with scale-like structure with a single stroke.
Although the system allows users to adjust the position and orientation of the scales, the options of the texture distribution is limited to only scale-like structure because the underlying method is based on a specific type of tesselation (i.e., Centroidal Voronoi) rather than general texture parameterization.
Tabby advances these works by enabling users to create general texture patterns for 3D models with complex geometries. 

\subsection{Autocomplete 2D Texture Creation}
Creating repetitive patterns is a tedious manual process.
To alleviate user workload, prior work has developed auto-completion techniques in 2D drawing applications
Kazi et al.~\cite{kazi2012vignette} demonstrated Vignette, an interactive drawing application that can facilitate user-defined 2D textures.
The user draws a part of a texture and gestures to automatically fill a 2D region with the texture.
Later research adapted this concept to enable data-driven decorative patterns~\cite{lu2014decobrush}, animated textures \cite{kazi2014draco}, and autocompletion of hand-drawn repetitions~\cite{xing2014autocomplete,xing2015autocomplete}.
These systems inspired us to investigate how we can integrate these autocomplete techniques into 3D modeling to support interactive tactile texture design.

\section{Design Goals}

We conducted a formative study to discover the needs and challenges in performing texture pattern creation with current 3D modeling tools.  
We observed the design process of two different types of users.
First, we participated in a workshop at a local community center where 25 novice users (most of them were high school students) designed 3D printed objects using entry-level CAD tools (TinkerCAD)
Next, we asked three professional Solidworks (CAD) users to apply a repeated texture to several complex geometries.
We conducted brief interviews with both user groups to learn from their experience.
These observations and interviews led us to the following high-level design goals.

\subsection{D1: Liberating from 3D Operations}
Prior work found that novice users often struggle to develop sufficient mental models in 3D modeling tools~\cite{hudson2016understanding}.
This causes difficulty in aligning or manipulating objects in 3D space.
Our study also confirmed that novice users struggle to precisely locate objects in a 3D space.
They needed to continuously change the camera angle to ensure that all texture elements were attached correctly to the target surface (Figure~\ref{fig:design-goals}A). 

This task becomes even harder on curved surfaces because the user must deform a texture to fit the target curvature (Figure~\ref{fig:design-goals}B).
Otherwise, the texture merely touches the surface at a single point.
However, this type of deformation task requires extensive 3D modeling skills.
In fact, our participants from the professional community confirmed that such a task often becomes intractable even for experienced users.
As a result, we observed that some participants from the experienced user group simply gave up creating textures on non-flat surfaces, or unwillingly accepted undesirable outcomes. 

These unsuccessful examples illustrate typical issues that CAD users encounter in 3D modeling.
Prior work found that 3D modeling becomes manageable even by novice users when it is presented as a series of 2D operations~\cite{hudson2016understanding}.
We thus should also exploit 2D operations in our system. 

%
%

\subsection{D2: Liberating from Repetitive Operations}
Tactile textures often consist of repeated patterns.
We observed that our participants from novice users attempted to achieve this through manually copying and pasting an element (Figure~\ref{fig:design-goals}C).
This na\"{i}ve approach can be feasible if the target surface is simple and flat.
But, it becomes intractable when the target geometry is more complex and curved because users must adjust the position and orientation of each element.
In addition, this simple copy-and-paste approach lacks the flexibility for design explorations.
After the texture elements are placed, if there is a need to iterate on the design, it would be very difficult to adjust the locations or orientations of all these elements manually. 

Some advanced CAD tools, such as SolidWorks or OpenSCAD, allow users to define and control the parameters of repetitive patterns to automate this process.
However, the parametric approach sometimes fails in non-symmetric geometries.
We asked professional SolidWorks users about the strategy to apply parameteric design for complex surfaces. 
According to their responses, the tool typically supports only symmetric shapes in automatically repeating the texture, so they must perform combinations of symmetric operations to approximately cover the surface (e.g., line patterns for a flat surface and circular patterns for a rounded surface).
Such tasks can be very difficult, and some of the professional users confessed it can be easily intractable to automate texture creation for complex surfaces.

Our tool, therefore, should reduce users' manual operations, even when interacting with complex surfaces.
We observed that users typically perform copy-and-paste operations to create repetitive patterns; thus, a repetition of such operations can signal an attempt to create tactile textures.
Our system should exploit this behavior to infer the user's intentions and, if possible, automatically complete the intended texture. 

\begin{figure*}[t!]
\centering
\includegraphics[width=2.1\columnwidth]{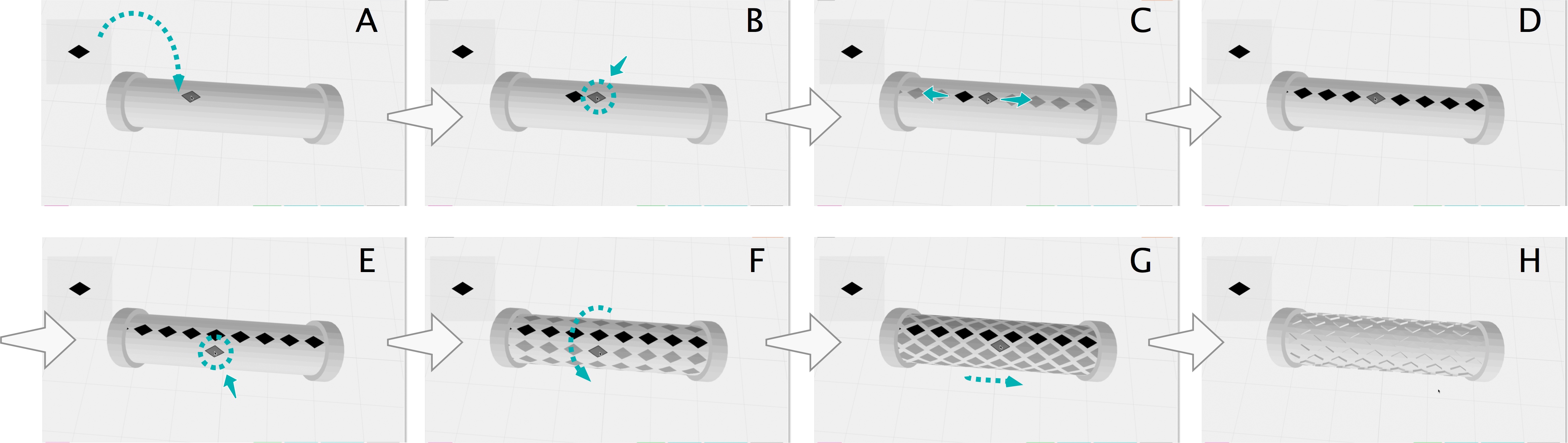}
\captionof{figure}{An overview of the workflow for creating gripping surface texture with Tabby. (A) Users sketch a texture element for grips and drag-and-drop it onto the 3D model. The system infers a possible target surface based on the user interaction. (B) Users copy and paste the texture element a couple of times. (C) When the system detects repetitive operations, Tabby suggests texture patterns. (D) Users can accept the suggested pattern. (E -- G) Another example of texture creation with Tabby. (H) Users extrude the texture, creating a 3D-printable pattern.}
\label{fig:workflow}
\end{figure*}

\subsection{D3: Allowing Intuitive Control}

With our interview, we found that designers typically want to control the following design elements:
\begin{enumerate} 
\item type of texture (e.g., bump, embossed, and hollow);
\item design of the texture unit (e.g., shape, size, and rotation);
\item alignment of the texture pattern (e.g.,  along a line, on a grid, or at random);
\item tactile properties (e.g., roughness and smoothness); and
\item region to fill the texture (e.g., roof of a house, a specific area in a tactile map).
\end{enumerate}

Although experienced designers can control these properties with  parameter tuning, such design process leaves them with a large gulf of evaluation~\cite{norman1986user}.
For example, one participant of our formative study suggested that he sometimes attempts to randomly change parameter values in an effort to achieve a certain effect because he does not fully understand the association between parameter values and their effects on the design.
Moreover, it could be even more difficult when designers have to deal with multiple parameters and constraints simultaneously.
Therefore, one design goal of our tool is to bridge this gulf by supporting \emph{direct manipulation} such that designers can more easily map their actions to intended results.

The ability to control the shape and patterns of texture may not be enough.
In many scenarios, users may want to specify a certain region to add texture (Figure \ref{fig:design-goals}D).
For example, when users design the texture for an object, such as the sole of shoes, scales of fish, the grip of robotic arms, and the tread of tires, they often need to choose a certain region and add tactile textures only to that region, instead of over the entire object. Therefore, another design goal is to simplify region selection. For this goal, we incorporate a semantic region selection method (described later).

\section{Tabby: Interface and User Interaction}


This section introduces Tabby, a system to support the design process of rich, user-defined tactile texture creation.
Tabby's design environment resembles that of a typical existing 3D modeling tool.
Our user interface consists of two components: a 3D model viewer and a 2D sketching canvas to design texture elements. 

The typical workflow in Tabby is as follows:

\emph{Step 1:} Import a 3D object into Tabby's working space. 

\emph{Step 2:} Sketch or import an SVG image for a desired texture element, and place it on the object (Figure~\ref{fig:workflow}A). 

\emph{Step 3:} The system infers the target region based on the placement of the texture element. Users can accept the suggested pattern (Figure~\ref{fig:workflow}D) or adjust the region selection if needed. 

\emph{Step 4:} After users copy and paste the texture elements a couple of times, the system suggests auto-completion of the pattern  (Figure~\ref{fig:workflow}B, C, E and F). 

\emph{Step 5:} Users adjust the properties of the pattern (Figure~\ref{fig:workflow}G). 

\emph{Step 6:} Once users confirm, the system extrudes the 2D texture, converting it into a 3D geometry while maintaining water-tightness (Figure~\ref{fig:workflow}H). 

\emph{Step 7:} Users can download the modified model to 3D print. 

Tabby has the following key features that address issues identified in our formative study and make the system usable for realistic applications.

\subsection{Semantic Region Selection}
\begin{figure}[!ht]
\centering
\includegraphics[width=1\columnwidth]{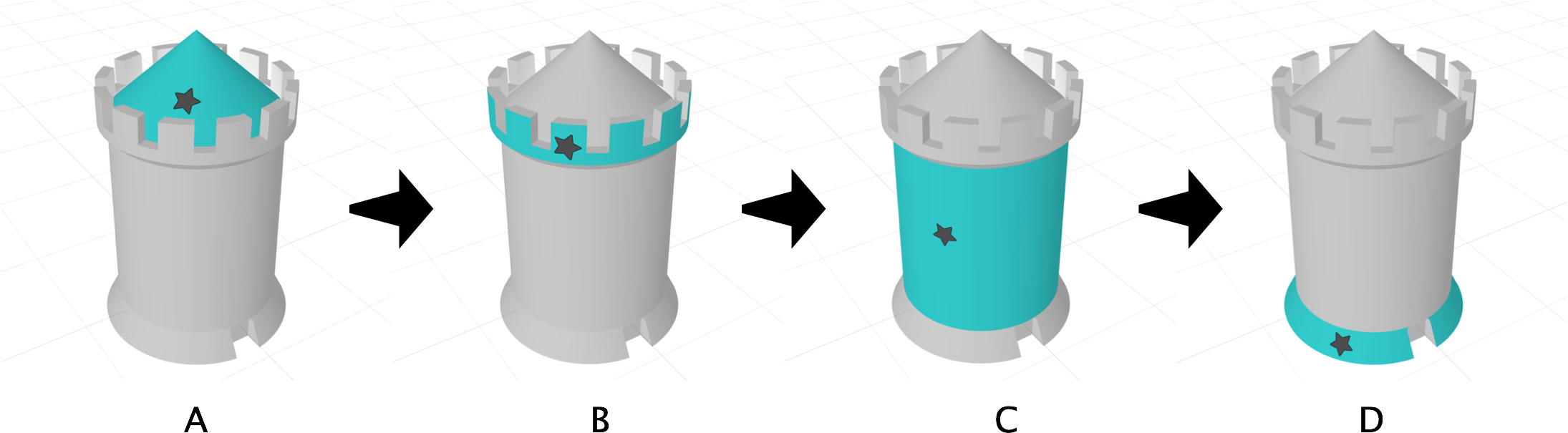}
  \caption{Automatic target region selection. Based on the current texture position, the system automatically infers on which surface she desires to create a tactile texture (highlighted in light blue).}~\label{fig:region}
\end{figure}

Users start with defining an element for texture patterns.
In Tabby, they draw the element in a 2D sketching canvas or import the element as an SVG file.
After deciding the element, users drag it into the main working space.
The system displays a shadow of the element as visual feedback.
As users move the element, the system automatically infers the surface region where they intend to create textures, and highlights it in light blue (Figure~\ref{fig:region}).
They may refine the system's suggested surface region by changing the region's size or by specifying the cutting boundaries with a mouse click.
When users release the mouse button, the system confirms the initial position of the element.
In this manner, Tabby liberates users from performing complex 3D operations (D1) and automates the task of selecting target regions (D3).

\subsection{Texture Auto-completion}
After users place the first element, they can perform copy-and-paste operations to start forming a texture.
Our formative study found that it is often very difficult to place new elements properly along the target surface.
Tabby supports this task by auto-completing repetitive patterns.

\begin{figure}[!h]
\centering
\includegraphics[width=1\columnwidth]{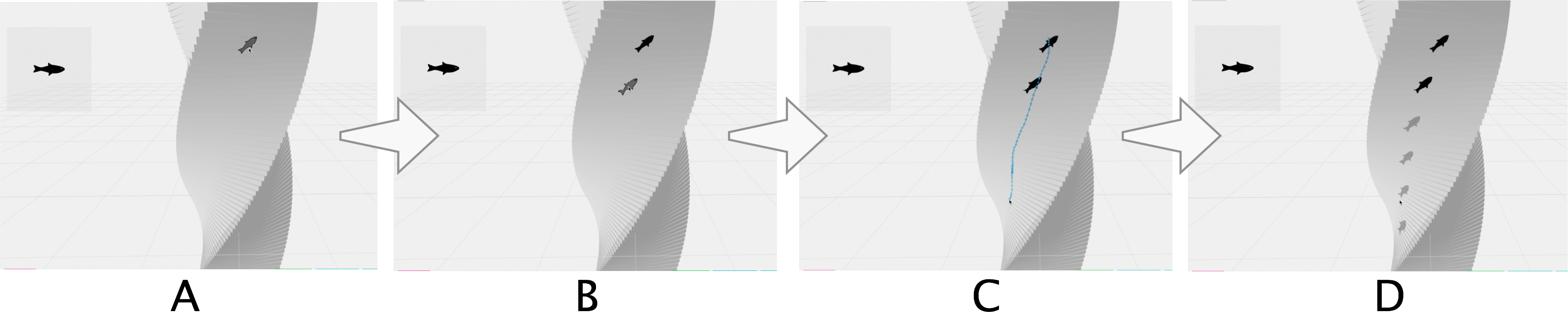}
  \caption{Auto-completion of repeated patterns by drawing a line. (A-B) Users copy and paste the element. (C) Users can demonstrate the line to create a pattern. (D) The system suggests a complete pattern along with the curved line.}~\label{fig:auto-complete}
\end{figure}

Tabby's auto-completion process is as follows. 
Once the system detects the user's copy-and-paste operations.
When such operations are detected, it tracks the placements of the first two texture units and calculates the relative positions between the two.
Then, Tabby makes a suggestion for auto-completion by presenting an example where each individual element is visualized as a shadow cast on the surface (Figure~\ref{fig:workflow} and ~\ref{fig:auto-complete}).
Tabby's extrapolation can support both patterns in x and y, and curves lines with the user's additional demonstration. 

Tabby can also support flexible customization.
When the user changes the position of the current element, the system shows the real-time preview to demonstrate how the rest of pattern can be changed.
With this preview, the user can adjust the properties of the repeated pattern, such as the density, the element size, and the rotation.  
In this manner, Tabby liberates users from repetitive operations while preserving the intuitive design process (D2).

After completing the design, the system automatically converts the 2D drawing element into a series of triangle meshes to create 3D textures.
Users can also interactively change texture types (e.g., bumps or cutting holes). 
As we will show in the following section, the system also ensures that these added triangle meshes are properly fused into the target to obtain a water-tight result.

\section{Method}
In this section, we present algorithms we have developed in order to support Tabby's key interactive features. 
Tabby converts a 2D drawing into a 3D printable texture.
A naive method would perform a linear extrusion of a 2D shape to obtain a 3D solid of the same shape and copy it over the target surface.
But this works well only when the target surface is flat. 
In order to support general applications where a target surface may have an arbitrary curvature, we must address two technical challenges: (1) each texture element must be deformed individually in order to smoothly conform to the curved surface; and (2) the resulting object maintains water-tightness for 3D printing.

Our method consists of two key components: (1) it utilizes UV mapping to achieve the evenness in sizes and spatial arrangements across the target surface, (2) it triangulates 2D sketch onto the target surface to achieve smooth deformation for three different types of textures.
Here we present the details of our method.

\subsection{Texture Mesh Generation}

\begin{figure}[h!]
\centering
\includegraphics[width=1\columnwidth]{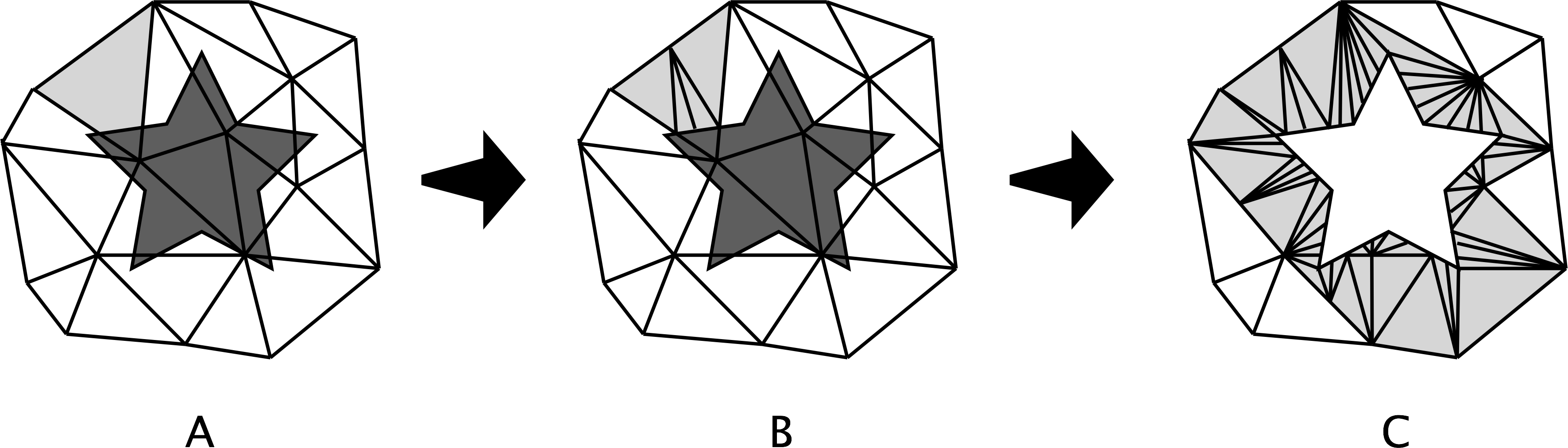}
  \caption{Texture mesh generation. (A) Obtain a diff region between a triangle and the texture. (B) Triangulate the diff region. (C) Replace the surfaces with the new one.}~\label{fig:mesh-generation}
\end{figure}

Let $M$ be the original mesh (the target to which a tactile texture is to be added) that consists of sets of triangle mesh surfaces $F$ and vertices $V$ in $\mathbb{R}^3$ ($|F| = N_F$ and $|V| = N_V$).
We compute a UV map for $F$ that gives us the UV coordinates of each vertex in $V$. 
We adopt the as-rigid-as-possible (ARAP) mesh parameterization \cite{liu2008local} to generate UV coordinates.
Next, we define $T_i$ as triangle points of UV mapping of the face $F_i$, and $T_i = [t_i^0, t_i^1, t_i^2] \ (t_i^j \in \mathbb{R}^2)$.
Let $P$ be a 2D shape (based on user's sketch) which consists of a set of 2D coordinates of stroke points.
For each $F_i \in F$, we calculate the overlapping region between $F_i$ and $P$ by performing boolean operations between $T_i$ and $P$ with Greiner–Hormann algorithm (Figure.~\ref{fig:mesh-generation} A).
We denote the result points as $P_i'$.
Once we obtain $P_i'$, then we triangulate $P_i'$ to get faces $F_i' = [f_{i, 1}', \cdots, f_{i, n}']$ that span the diff regions (Figure.~\ref{fig:mesh-generation} B).
To perform the triangulation, we use the standard constrained Delaunay triangulation~\cite{chew1989constrained}.
Through triangulating each face, we can obtain the boundary points in 2D coordinates and corresponding triangle faces.
Then, we replace the original mesh surface $F, V$ with new meshes $F', V'$, where $V'$ are 3D vertices converted from 2D coordinates of $P' = \{P_i'\}$ (Figure.~\ref{fig:mesh-generation} C). 
Algorithm~\ref{alg:texture-generation} illustrates this process in pseudo code.

\begin{algorithm}
\caption{Texture Mesh Generation}\label{alg:texture-generation}
\begin{algorithmic}[1]
\State \textbf{Input:} $V, F, P$
\State \textbf{Output:} $V', F'$
\\
\For{$i = 0, \cdots, Size(F)$}
\State $T_i \gets getVertexUV(F[i])$
\State $P_i \gets getDiffPoints(T_i, P)$
\If {$Size(P_i) > 3$}
  \State $removeFace(F[i])$
  \State $F_i' \gets triangulate(P_i)$
  \State $V_i' \gets getVertex3D(F_i', P_i)$
  \For{$j = 0, \cdots, Size(F_i')$}
    \State $addFace(F_i'[j], V_i'[j])$
  \EndFor
 \EndIf
\EndFor
\end{algorithmic}
\end{algorithm}

\subsection{Texture boundary extrusion}
\begin{figure}[!h]
\centering
\includegraphics[width=1\columnwidth]{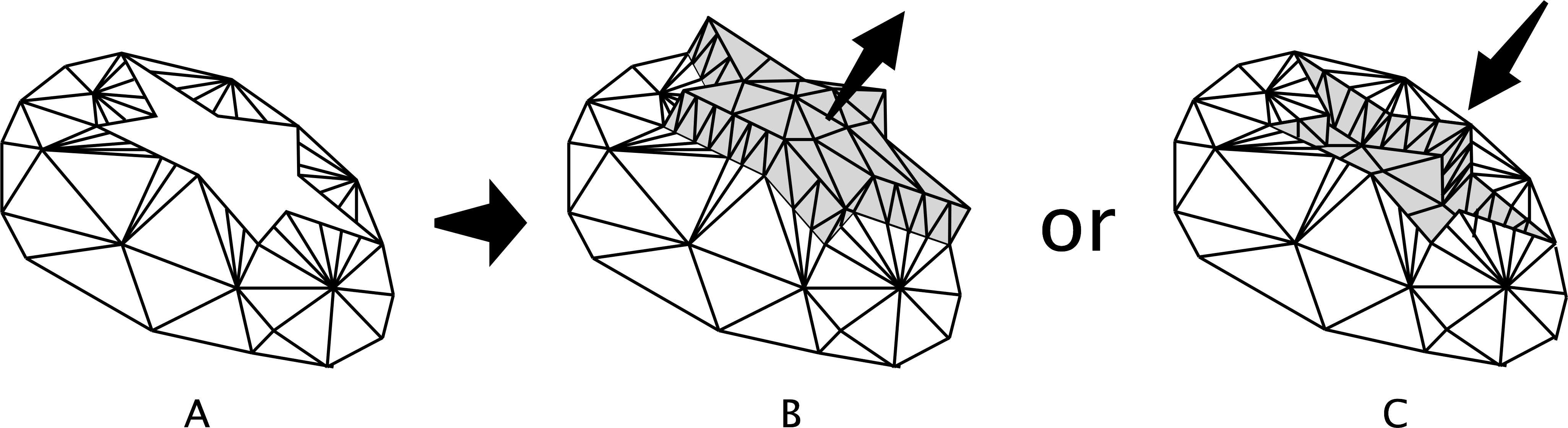}
  \caption{Extrusion operations. (A) After the system defines the boundary vertices, it computes the normal vector for each vertex. (B) When the system performs extrusion, it automatically adds walls around the texture element based on the obtained vertex normals. (C) With the same operation, the system can perform intrusion. }~\label{fig:mesh-extrusion}
\end{figure}

Once the boundary positions and surrounding surfaces are determined, the system creates a corresponding enclosure to maintain the water-tightness.
We compute the vertex normal vector for each boundary point, and get a set of outer vertices by extending with the normal vector.
We then perform the same constrained Delaunay triangulation on the outer vertices to obtain outer polygonal surfaces (Figure.~\ref{fig:mesh-extrusion} B).
Note that the resulting textures are smoothly deformed to fit the target curvature.
With the same interactive operation, users can also obtain an embossed texture by intruding the boundary points (Figure.~\ref{fig:mesh-extrusion} C) or a hollow texture by creating internal cavities inside the mesh with a consistent wall thickness.


\subsection{Interactive Mesh Segmentation}

Tabby enables the user to interactively select a semantic region.
To enable this, we leverage and extend the existing mesh segmentation algorithms.
Our method is based on cross-boundary mesh decomposition~\cite{zheng2010mesh}, which computes a harmonic field with a Laplacian matrix to obtain segmentation boundaries by cutting along an isoline of the harmonic field.
To obtain boundary positions, previous work requires users to explicitly specify the boundary points by user interaction, such as selecting the foreground or background regions~\cite{ji2006easy}, drawing strokes along the boundaries~\cite{guy2014simselect}, and clicking a point near the boundaries~\cite{zheng2012dot}.
However, these specifications of cutting boundaries require additional user interaction for mesh segmentation.
Thus, we extend these methods to automatically infer the surrounding segmented region based on the current mouse position.

To achieve this goal, we obtain the cutting boundaries by computing distortion metrics.
We hypothesize that highly distorted vertices can be potential candidates of boundary points because the distortion occurs in high surface curvature, and cutting boundaries are also most likely located in high curvature points.
To compute the degree of distortion, we use the Gaussian curvature of the vertices across a larger surrounding region~\cite{sheffer2002seamster}.
The distortion of a vertex $i$ is defined as: 
\begin{eqnarray*}
D (i) = \max_{0 \le r \le R} \frac{2 \pi - \sum_{j} \tau_j (r) }{2 \pi}
\end{eqnarray*}
where $R$ is a region radius and $\tau_j (r)$ are the angles at $i$ of face $j$ inside of region radius $r$.
After calculating the distortion of each vertex, we extract high distorted points using a terminal vertex selection algorithm~\cite{sheffer2002seamster} with weights defined by the distance between the current mouse position and the target vertices.

\begin{figure}[!h]
\centering
\includegraphics[width=1\columnwidth]{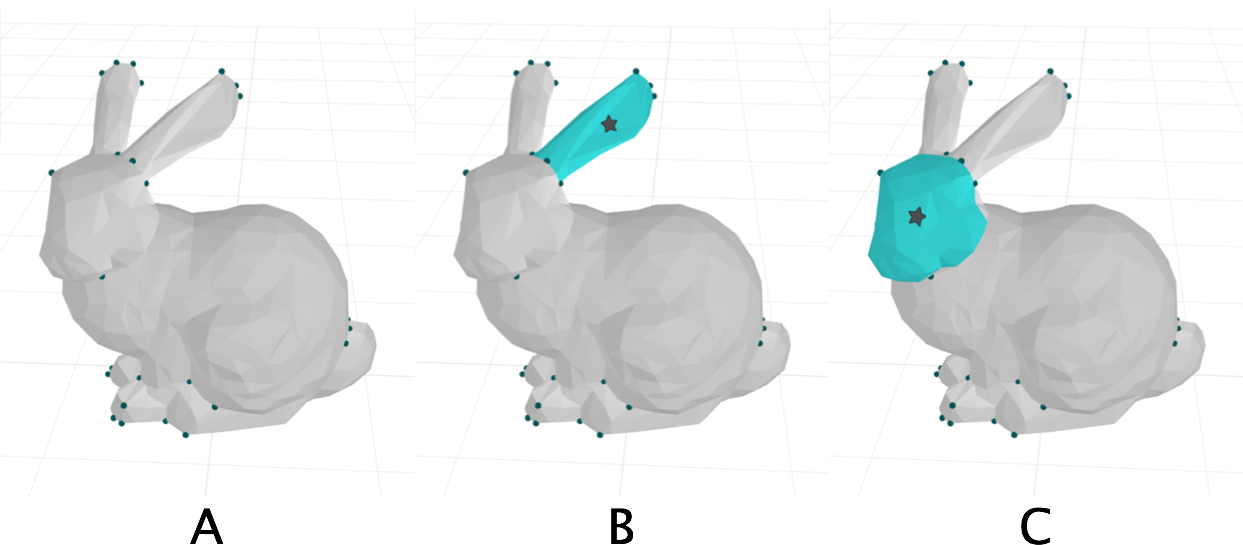}
  \caption{Mesh segmentation. (A) The system identifies high distortion points (dots in the model). (B) Tabby uses these points as boundary candidates, and segments surface regions. (C) Based on the cursor position, the system updates harmonic fields and suggests possible target regions.}~\label{fig:mesh-segmentation}
\end{figure}

Figure~\ref{fig:mesh-segmentation} illustrates the high distortion points around the current mouse position.
Once we obtain the boundary candidates with the distorted points, we compute a set of harmonic fields that propagate in different directions and select the best isolines.

\begin{algorithm}
\caption{Mesh Segmentation with Harmonic Field}\label{alg:2}
\begin{algorithmic}[1]
\State \textbf{Input:} $V, F, L$
\State \textbf{Output:} $Isolines$
\\
\For{$i = 0, \cdots, Size(V)$}
\State $D[i] \gets computeDistortion(V[i])$
\EndFor
\State $Bnd \gets getBoundaries(D)$
\For{$i = 0, \cdots, Size(Bnd)$}
\State $v_1, v_2 \gets getNeighborPoints(Bnd[i])$
\State $V_1.push(v_1)$
\State $V_2.push(v_2)$
\EndFor
\State $P, B \gets getPositionalMatrix(V_1, V_2)$
\State $phi \gets computeHarmonicField(L, P, B)$
\State $computeIsolines(phi)$
\end{algorithmic}
\end{algorithm}

\section{User Evaluation}
To validate the usability of Tabby, we conducted a controlled experiment.
The objective of this evaluation was a performance comparison in texture creation and modification tasks between Tabby and commercial CAD tools.

\subsection{Participants}
We recruited 7 participants (5: male, 2: female) who major in various fields (computer science, mechanical engineering, architecture, and industrial design) in a local university.
Tabby is designed to support both novice and experienced users, but we expect the non-experience users may not be able to complete tasks in the reference system.
Therefore, to make a comparative study feasible, we targeted intermediate- or expert-level CAD users in this experiment.
The participants had at least 10-month experience in using commercial CAD tools (Min: 10 months, Max: more than 6 years, Average: 3.2 years).
The tools they chose include both novice-oriented (e.g., TinkerCAD and 123D Design) and expert-oriented (e.g., SolidWorks and Rhinoceros) traditional CAD software.
They self-reported the expertise of 3D modeling tools as 5.3 out of 7 in average.

\subsection{Tasks}
We asked participants to add and edit a tactile texture on existing models.
The task consisted of two different operations: creating and modifying textures.
We prepared six different basic geometries as target surfaces.
We chose six target surfaces to cover a variety of complexity: 1) a flat rectangle; 2) the side of a cylinder; 3) a sphere; 4) the side of a cone; 5) the head of a knight \footnote{\url{http://www.thingiverse.com/thing:1301815}}; and 6) the body of the Stanford bunny.
In the creation tasks, the participants were instructed to create a texture onto the given surface by placing a small cylinder texture element in a 3x3 grid.
In the modification tasks, we asked participants to change the size of the texture elements which they had created in the previous creation task. We measure the completion time of two geometries: a sphere and cone.

\subsection{Procedure}
Each participant first completed the tasks with the existing tools in their computers.
We allowed them to choose a tool they felt most comfortable with using.
Four participants chose SolidWorks, one chose Rhinoceros, and two participants chose TinkerCAD.
They were asked to create textures on all target surfaces except the last.
We excluded the last condition (the Stanford bunny) because we expected that it would be highly unlikely to be completed in five minutes.
We also allowed the participants to give up after five minutes if they felt that they would not be able to complete a task.

After completing all the tasks with their tools, they were instructed to perform the same tasks with our tool.
We gave participants a brief instruction about Tabby, and provided time for familiarizing themselves with the system.
We recorded their interactions and measured their completion time in all conditions.
At the end of the experiment, we conducted a semi-structured interview where we asked their experiences with Tabby, opinions about its benefits and drawbacks, and potential use cases.
Each participant was offered a \$15  Amazon gift certificate as a compensation at the end of the experiment.


\section{Results}
\begin{figure}[!h]
\centering
\includegraphics[width=1\columnwidth]{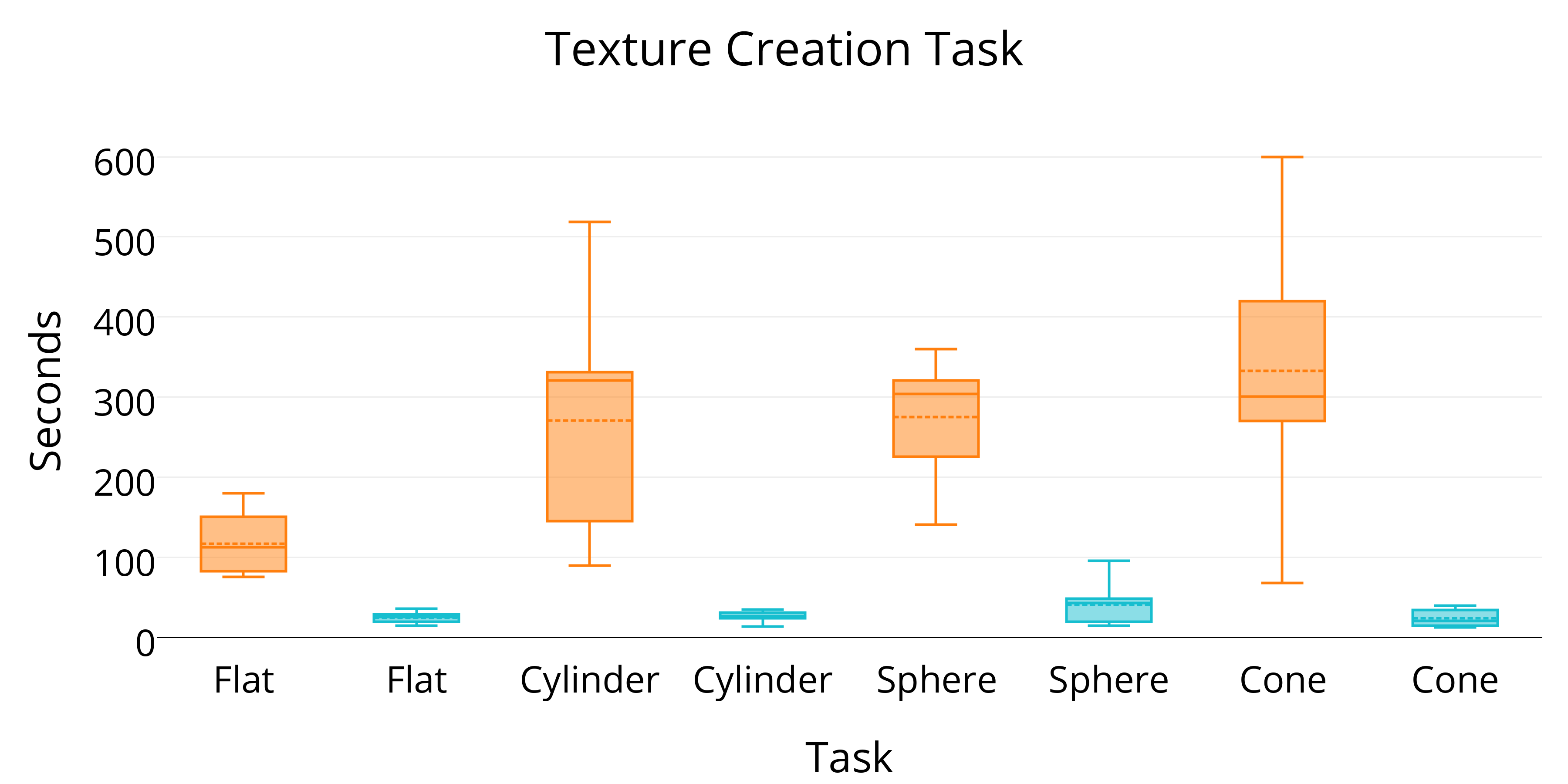}
  \caption{Completion time of the texture creation tasks with the reference tool (orange) and Tabby (light blue). Each box represents the median (the solid line), the mean (the dotted line), and the quartiles (the top and bottom edges of the box) in each condition.}~\label{fig:results}
\end{figure}

Figure~\ref{fig:results} illustrates the overview of the completion time for each texture creation task.
With Tabby, participants completed their tasks with 29 seconds on average (Flat: 24 sec, Cylinder: 26 sec, Sphere: 41 sec, Cone: 24 sec).
In contrast, in the reference condition, they needed 248 seconds on average (Flat: 117 sec, Cylinder: 270 sec, Sphere: 275 sec, Cone: 332 sec).
Two participants even gave up completing the tasks for a sphere and cone surfaces.

A quantitative analysis of the task completion time reveals that the Tabby substantially reduced the time to perform texture creation tasks.
Our repeated-measure ANOVA test found significant results for both tools  ($F_{1,6}=48.6$, $p<.01$) and surfaces ($F_{3,18}=6.19$, $p<.01$) as well as their interaction ($F_{3,18}=6.03$, $p<.01$).
Our post-hoc pairwise comparison reveals significant differences between Tabby and the reference tools.
It also confirmed significant differences in the tasks; the flat surface was significant faster than the others.

We further examined the effect of the interaction of the tools and surfaces.
A repeated-measure ANOVA test for the performance time revealed that the surfaces were a significant factor with the reference systems ($F_{3,18}=6.18$, $p<.01$) but not with Tabby ($F_{3,18}=2.39$, $p=.10$).
This suggests that surface complexity greatly influenced on performance with the reference systems while the completion time was more comparable with Tabby.




\begin{figure}[!h]
\centering
\includegraphics[width=1\columnwidth]{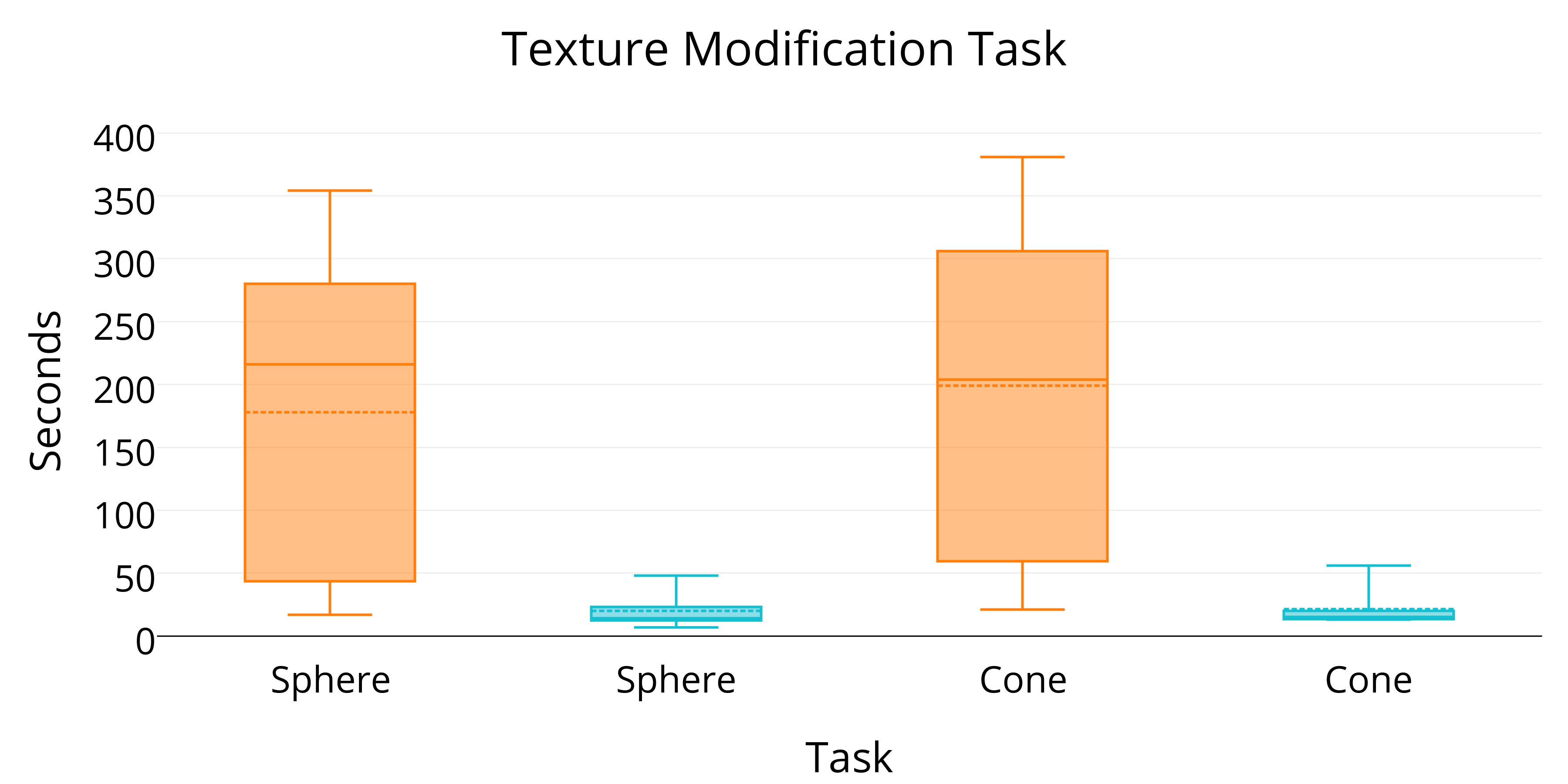}
  \caption{Completion time of the modification tasks with the reference tool (orange), and Tabby'(light blue).}~\label{fig:results-modification}
\end{figure}

Figure~\ref{fig:results-modification} depicts the result in the modification tasks.
Our two-way repeated-measure ANOVA found a significant effect in tools ($F_{1,6}=12.7$, $p<.05$) but not in the two surfaces ($F_{1,6}=0.72$, $p=.42$).
A post-hoc test confirmed that Tabby was significantly faster than the reference tools.


On 7-point Likert scale questions (1: strongly disagree -- 7: strongly agree), participants rated our tool as useful (6.3),  easy to use (5.7), and effective to perform the task (6.4).
The participants liked each feature of the Tabby useful (texture generation: 6.7, auto-completion: 6.1, drag-and-drop manipulation: 5.5, and sketching: 5.7).

\begin{figure*}[t!]
\centering
\includegraphics[width=2.1\columnwidth]{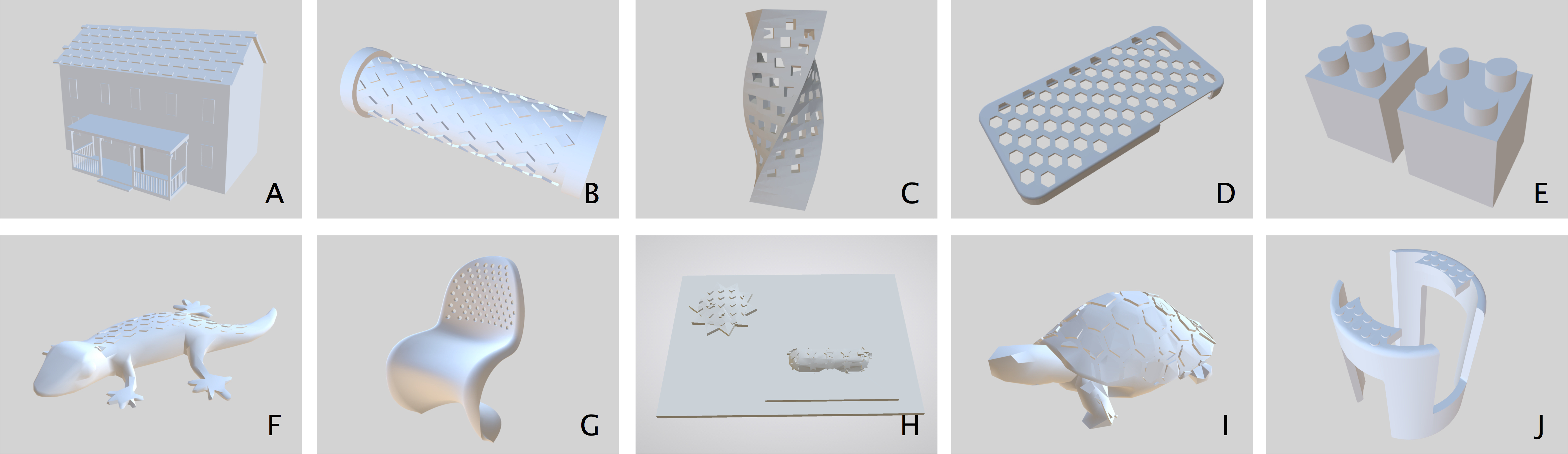}
  \caption{Examples of possible real-world use scenarios. All models are created within one minute design time by the authors. (A) A roof tile pattern on an architectural model. (B) A grip pattern on a bike handle. (C) A light design for lampshades. (D) A repeated pattern for smartphone cases. (E) Stacking functionality for toy blocks. (F) Scale of geckos. (G) A design for chairs. (H) Tactile picture books. (I) Shell of turtles. (J) Anti-slip finishing for cup holders.}~\label{fig:applications}
\end{figure*}

\subsubsection{Simple and effective interaction}
Our user interface facilitates the intuitive operations of texture creation.
One participant commented that the simplicity of the interface helped them quickly adapt to the system.
\begin{quote}
P2: \textit{``I really like the simplicity of it. [...] This one is actually pretty easy, just by messing around this, you can really get a better understanding of how it works pretty quickly, compared to the SolidWorks which is extremely intimidating with a lot of buttons that doesn’t make sense to us.''}
\end{quote}

Participants also agreed that the user interface is simple yet powerful to effectively perform the task.
\begin{quote}
P4: \textit{``Also, the pattern generation tool was very helpful because instead of clicking and dragging like 10 different times, I only needed to copy or paste once for the full pattern to be generated.''}
\end{quote}

The auto-completion feature significantly reduced their workload.
P3 also appreciated the flexibility of the generated texture pattern.
\begin{quote}
P3: \textit{``You get one axis when you copy and paste in one direction and another axis you copy and paste in another direction. I like that there are a lot of options of how you can copy and paste as to how you get your pattern to wrap around the object is''}
\end{quote}

\subsubsection{Avoiding confusion in 3D manipulation}
With Tabby, participants successfully reduced the number of 3D object manipulation.
As Tabby allows users to perform texture design as a series of 2D operations, participants did not get confused about object manipulation in the 3D space.
\begin{quote}
P4: \textit{``I like that [in our tool] the pattern is automatically attached to the shape, because I don't have to make sure that there is no space between it. [...] TinkerCAD is hard to know where you're in the space, when clicking and dragging sometimes you think the object is sitting on the top of object, but when you rotate the workspace, it turned out it’s floating somewhere random in space.''}
\end{quote}

\subsubsection{Most effective for arbitrary geometries}
Participants liked that Tabby can support arbitrary geometries and automatically deform surface textures.
They agreed that wrapping a pattern around a complex surface becomes much faster [P6, P7] and easier [P5] than manual efforts.
P6 realized that if he tried to add textures on arbitrary curvatures, it took much more time than he initially thought:
\begin{quote}
P6: \textit{``I don't know if I could. I could do that, but it would take a while. I could not automate, I would have to manually make every texture element. I could copy and paste them, but thing is when I extrude it, from the sketch I would have to make different extrusion for each texture.''}
\end{quote}

For users who have less experience, even simple geometries like a cone and a sphere are difficult and tedious to apply the texture to.
To create repetitive patterns with the parametric modeling tools like SolidWorks, users have to define the plane and control the parameter over it.
However, this can easily become intractable once the target geometry becomes non-flat.
Because of this, two participants gave up the tasks in creating textures on non-flat curvature.
P2, one of these participants, expressed why he gave up the task:
\begin{quote}
P2: \textit{``I know that there is a way to do it, but I know the task is tedious. [...] When you're doing stuffs on rough surfaces like this [bunny], it takes forever to look more natural.''}
\end{quote}


Our user study results showed that Tabby was significantly faster than existing 3D tools across surfaces with different complexity levels.
With Tabby, participants completed a texture creation task almost five times faster even in the simplest target surface than commercial 3D modeling software.
Performance with Tabby was consistent whereas participants needed more time in the reference condition as target surfaces became more complex.
Our qualitative evidence also supported this quantitative result.
These results validate that Tabby meets its three design goals.

\section{Applications}

Our participants were able to articulate a variety of specific use scenarios for Tabby.
Examples are making gears [P1, P7], grips for robotic arms [P3], architectural models [P2], treads of a tire [P3, P6], wavy texture on a phone holder [P2, P6], scales of animals [P4], and tactile graphics and physical visualization [P4].

Based on their feedback, we later investigate whether our tool is capable of creating such possible applications.
Figure~\ref{fig:applications} demonstrates that our tool can potentially support a wide variety of use scenarios.
Tabby enables users to quickly apply texture for increasing the aesthetics of the design (lampshades, design of chairs, smartphone cases), adding the details for realistic models (architectural models, scales of animals), increasing the usability and accessibility (grips, tactile picture books for people with visual impairments), and enhancing the functionality (blocks).

\section{Limitations and Future Work}

Tabby has several technical limitations that require future work.
One major limitation is that our techniques heavily rely on 2D coordinates of UV mapping.
If the UV coordinates do not properly parameterize in surrounding areas, the resulting texture can be distorted.
For example, we noticed that sphere topology may produce distorted patterns.
One possible solution is to compute parameterization locally on a segmented area.
Because the system already identifies the segmented region, we may compute and update UV coordinates on the selected region~\cite{schmidt2006interactive}.
In addition, the discontinuity on a seam could cause problem associated with the UV mapping.
Currently, the system automatically cuts the shapes on the seam of the UV mapping.
But it may produce undesirable results, such as a discontinuous pattern on a cylinder surface.
Interestingly, our participants reported resulting patterns (e.g., distorted patterns on a cone or an edge of a roof tile or a grip) generally looked more natural (P2, P3) and organic (P4) than manual or parametric placement.
We will further need to investigate what aspect may affect the undesirable outcomes.

Another limitation is the lack of 3D texture geometry. 
3D printable textures have a rich design space of more complex geometry, such as hemisphere bumps or spikes, instead of simply extruding or intruding 2D sketches. 
One typical way to create such 3D textures is to use the displacement map discussed in~\cite{torres2015hapticprint}.
However, there are several trade-offs between our approach and the displacement map.
First, our approach can generate a hollow structure (e.g. a lampshade, a chair, and an iPhone case in Figure.~\ref{fig:cover}) which can be potentially difficult with the displacement map.
Another trade-off is that the displacement map typically requires a high-density mesh to produce a smooth surface.
For example, it requires 40,000 (200 x 200) vertices to generate 7 x 7 texture patterns~\cite{torres2015hapticprint}.
For a low-density mesh, the system may require the re-meshing process to increase the density of the vertices. 
To address these problems, we will seek to combine the displacement map with our mesh generation method to generate textures that have more complex geometry while maintaining the auto-completion workflow.

Our current implementation only supports a limited variety of texture patterns.
Such variations can be greatly improved using a data-driven approach to learn new patterns from examples.
For example, recent advances in 2D drawing texture creation have demonstrated that a data-driven algorithm can enable rich and nuanced expressions of possible designs~\cite{guerrero2016patex,lu2014decobrush,xing2014autocomplete}. 
Since our system exploits 2D drawing for 3D pattern creation, we expect it could be relatively easy to incorporate with these algorithms to enhance the variation of patterns.
As future work, we will extend the variation of the patterns by adopting these techniques.

\section{Discussions}

We also acknowledge that our evaluation study lacks a comparison with other state-of-the-art tools and recent research systems.
For example, tools like MeshMixer and Grasshopper also provide some form of rapid patterning functionality via menu-based commands. 
While our participants did not use these tools before, they were able to notice Tabby's comparative advantages.
They pointed out that the existing tools require them to know what commands to use and what effects to occur.
As the tool are more capable, such commands also increase and become more complex. 
However, most of these commands are simply not utilized by common users. 
For example, among the thousands of commands available in AutoCAD, typical users only utilize 31-40 of these commands~\cite{matejka2009communitycommands}.
On the other hand, they appreciate Tabby's ``situated'' interactive interface, which triggers a certain command (e.g., repeat a pattern) with an intuitive interaction (e.g., copy-and-paste) because they do not need to explicitly learn the commands beforehand.
We envision that our work is a stepping stone to redesign such command-based interface into the situated interface in the domain of 3D modeling software.

\section{Conclusion}

We present Tabby, an interactive tool to support designing and applying 3D printable textures on an arbitrary complex surface of the existing object.
To minimize repetitive manual efforts, we adopt the \emph{auto-completion} metaphor, which automatically infers the user's demonstration and suggests the possible desired patterns.
To enable this, we develop a series of techniques which infer the user's intention, select the semantic region, and convert 2D shapes into 3D textures.
Our controlled experiment shows that Tabby enables the participants to create and modify 3D texture much faster than conventional tools, especially for complex target surfaces.


\balance

\bibliographystyle{acm-sigchi}
\bibliography{references}

\end{document}